\input{aipcheck}

\documentclass[
    ,final            
  ]
  {aipproc}

\layoutstyle{6x9}
\newcommand{\hi}{H{\sc i }}

\begin{document}

\title{Early-type galaxies with neutral hydrogen in the Virgo cluster from
the ALFALFA survey}


\author{S. di Serego Alighieri}{
  address={INAF - Osservatorio Astrofisico di Arcetri, Firenze, Italy}
}

\author{M. Grossi}{
  address={INAF - Osservatorio Astrofisico di Arcetri, Firenze, Italy}
}

\author{C. Giovanardi}{
  address={INAF - Osservatorio Astrofisico di Arcetri, Firenze, Italy}
}

\author{S. Pellegrini}{
  address={Universit\`a di Bologna, Bologna, Italy}
}

\author{G. Trinchieri}{
  address={INAF - Osservatorio Astronomico di Brera, Milano, Italy}
}

\begin{abstract}
We extend our published work on the neutral hydrogen content of early-type
galaxies in the Virgo cluster using the catalogue of detected sources from
the ALFALFA survey, by showing the 21cm spectra of all the detected
galaxies and discussing a deeper analysis of the ALFALFA datacubes,
searching for lower S/N sources. A view
of the multiphase interstellar medium of M86 is also presented, by comparing
images of the cold, warm and hot phases.
\end{abstract}

\maketitle

\section{Neutral hydrogen in the Virgo ETG}

The multi-phase properties of the ISM of early-type galaxies (ETG) in a rich
environment, like the Virgo cluster, can provide important information on the
influence of interactions (merging, ram-pressure stripping, tidal interactions 
etc.) on galaxy evolution. The cold phase, of which neutral hydrogen is a major
component, has a hard time in surviving in the cluster environment, because of
the evaporation induced by the hot ICM and ISM, particularly in the massive ETG, where
hot gas is denser. Therefore \hi survival in cluster ETG may be the sign of
interesting phenomena. Di Serego Alighieri et al. (2007) have recently completed 
a survey of the \hi content in the Virgo cluster ETG, using the catalog of \hi 
sources produced by the ALFALFA survey (Giovanelli et al. 2007). They find that only a very small
percentage (2.3\%) of the ETG with $B_T \leq 18.0$ in the Virgo Cluster
Catalogue (VCC, Binggeli et al. 1985) have neutral hydrogen
above about $3\times 10^7 M_{\odot}$.
Most of the detected ETG are dwarf galaxies, which are at the border of the
early-type classification.

Figure 1 shows the \hi spectra of all the Virgo ETG detected in the
published ALFALFA catalog, except for VCC 2062, whose \hi line is almost
unresolved from the one of the gas rich spiral NGC 4694.
For the published catalog the ALFALFA team has adopted a S/N
threshold of 6.5 to discriminate between confirmed and spurious detections
in the blind search of the survey data-cubes (Saintonge et al. 2007).
To search for additional 21-cm sources at lower S/N ratios, we have
inspected the survey data-cubes at the optical position and
velocity (when available) of each object in our optical sample of Virgo
ETG obtained from the VCC. We did not find
any additional ETG convincingly detected in \hi.
The tentative faint \hi sources, which we identified at S/N$<$5, would need deeper
pointed follow-up observations to be confirmed. This negative result
reinforces our confidence in the completeness of the published ALFALFA catalog.

\begin{figure}[!ht]
\includegraphics[width=0.24\textwidth]{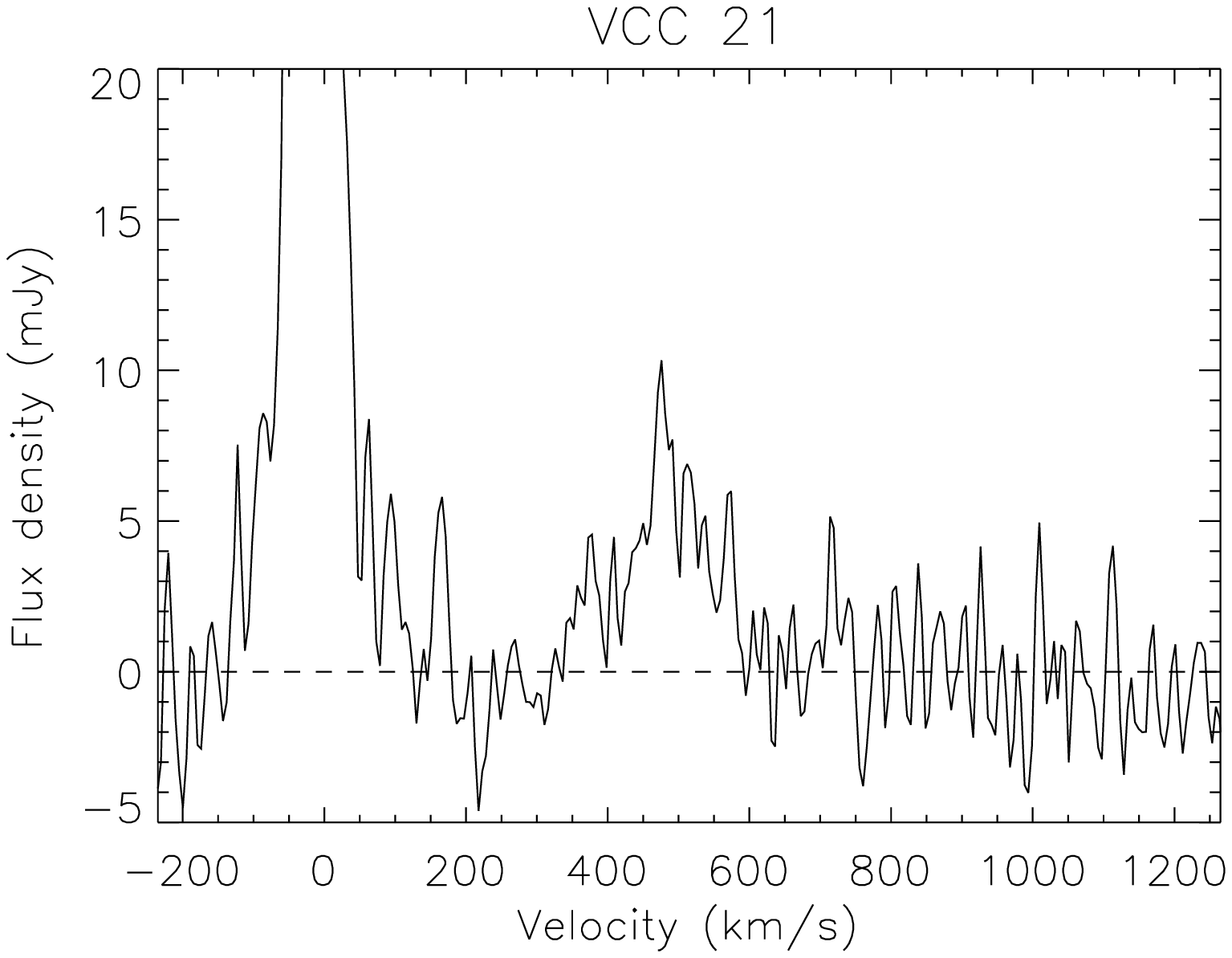}
\includegraphics[width=0.24\textwidth]{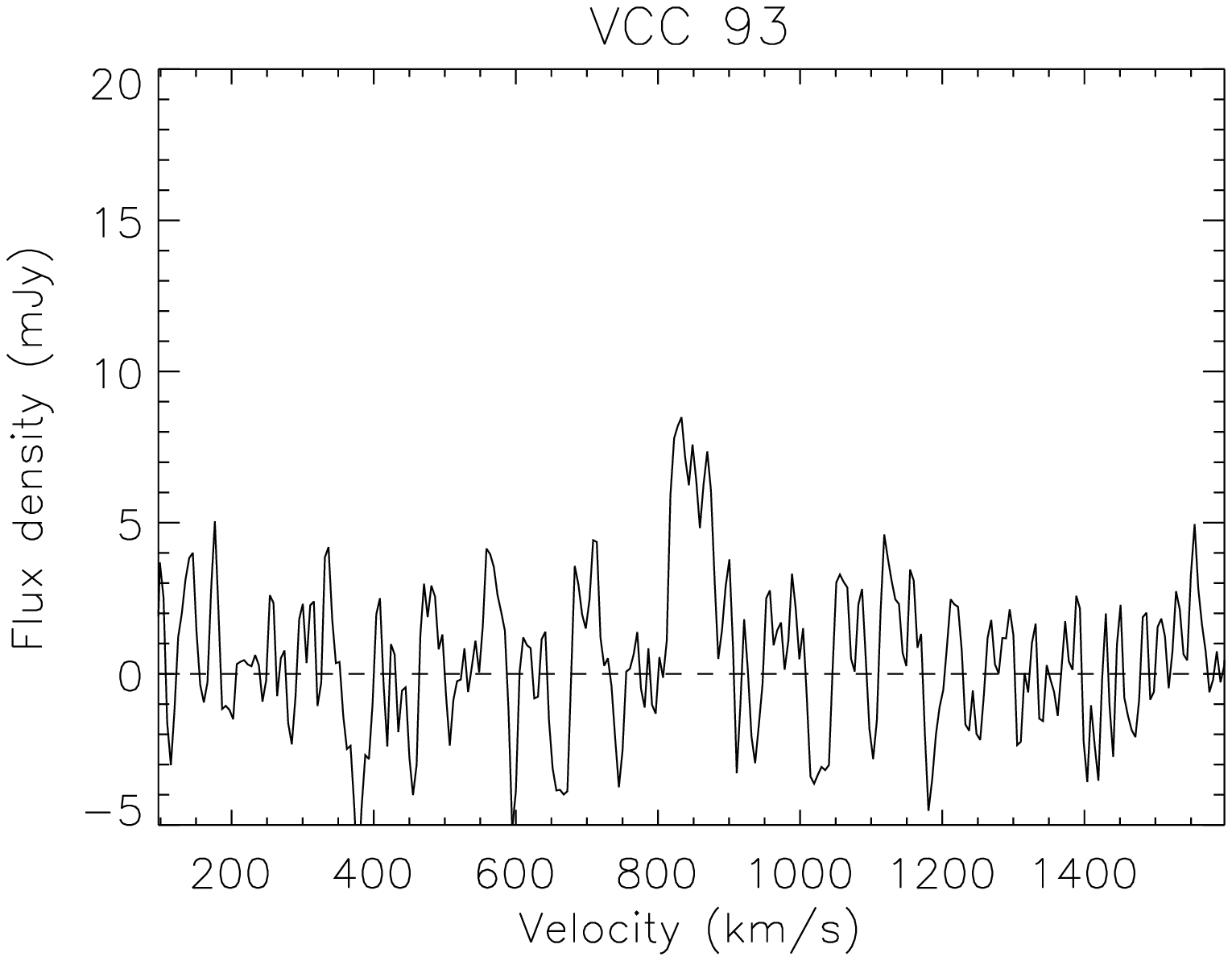}
\includegraphics[width=0.24\textwidth]{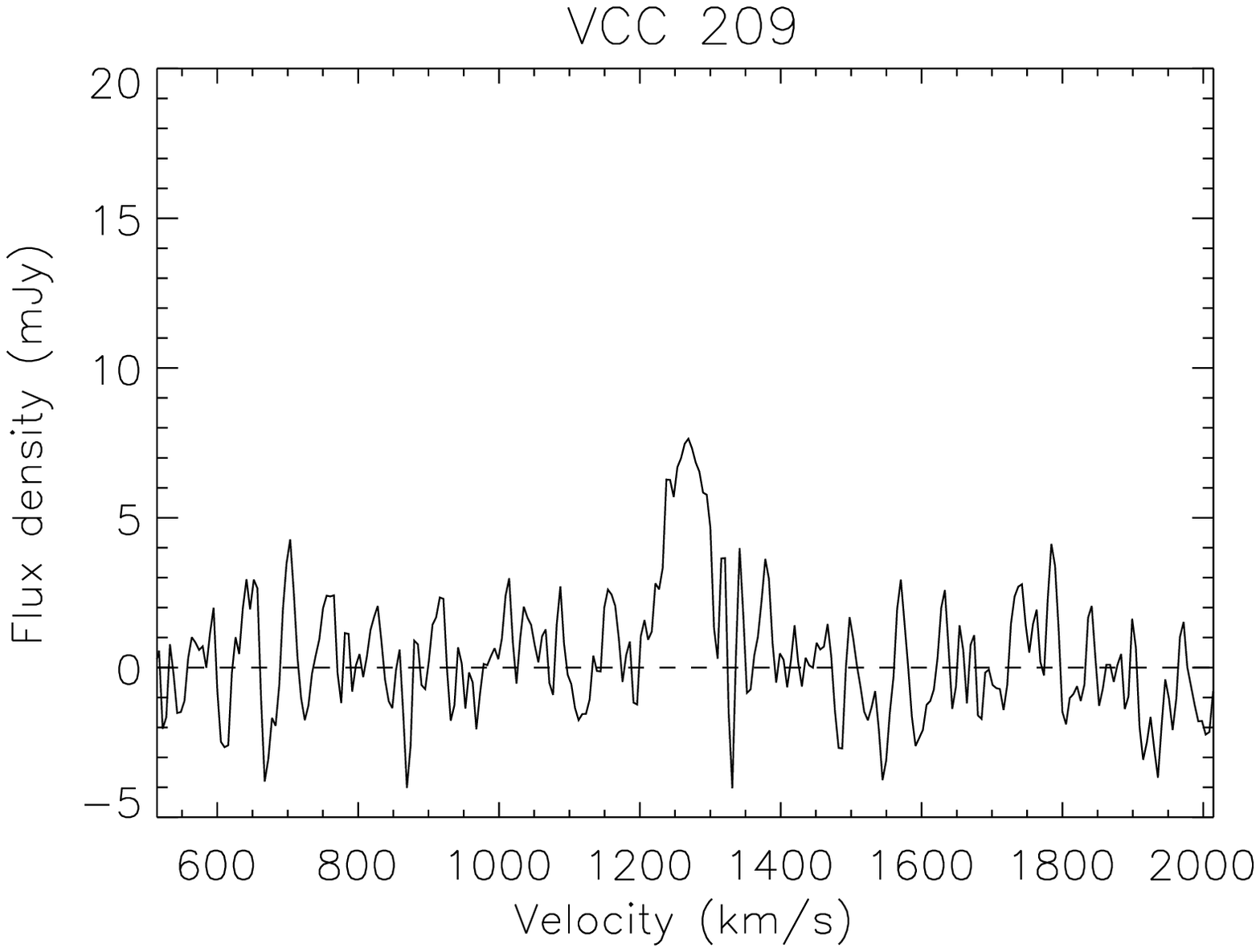}
\includegraphics[width=0.24\textwidth]{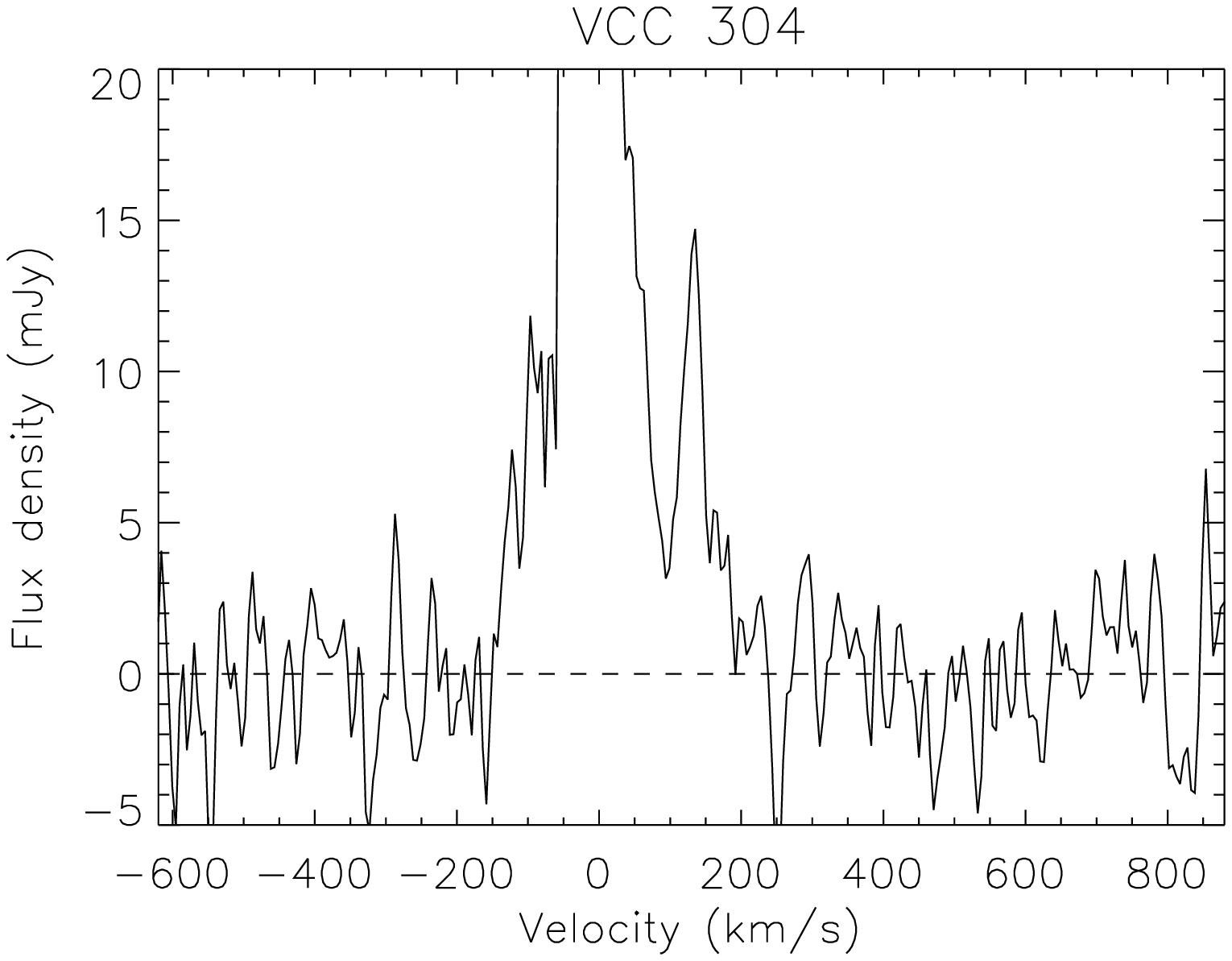}
\end{figure}
\begin{figure}[!ht]
\includegraphics[width=0.24\textwidth]{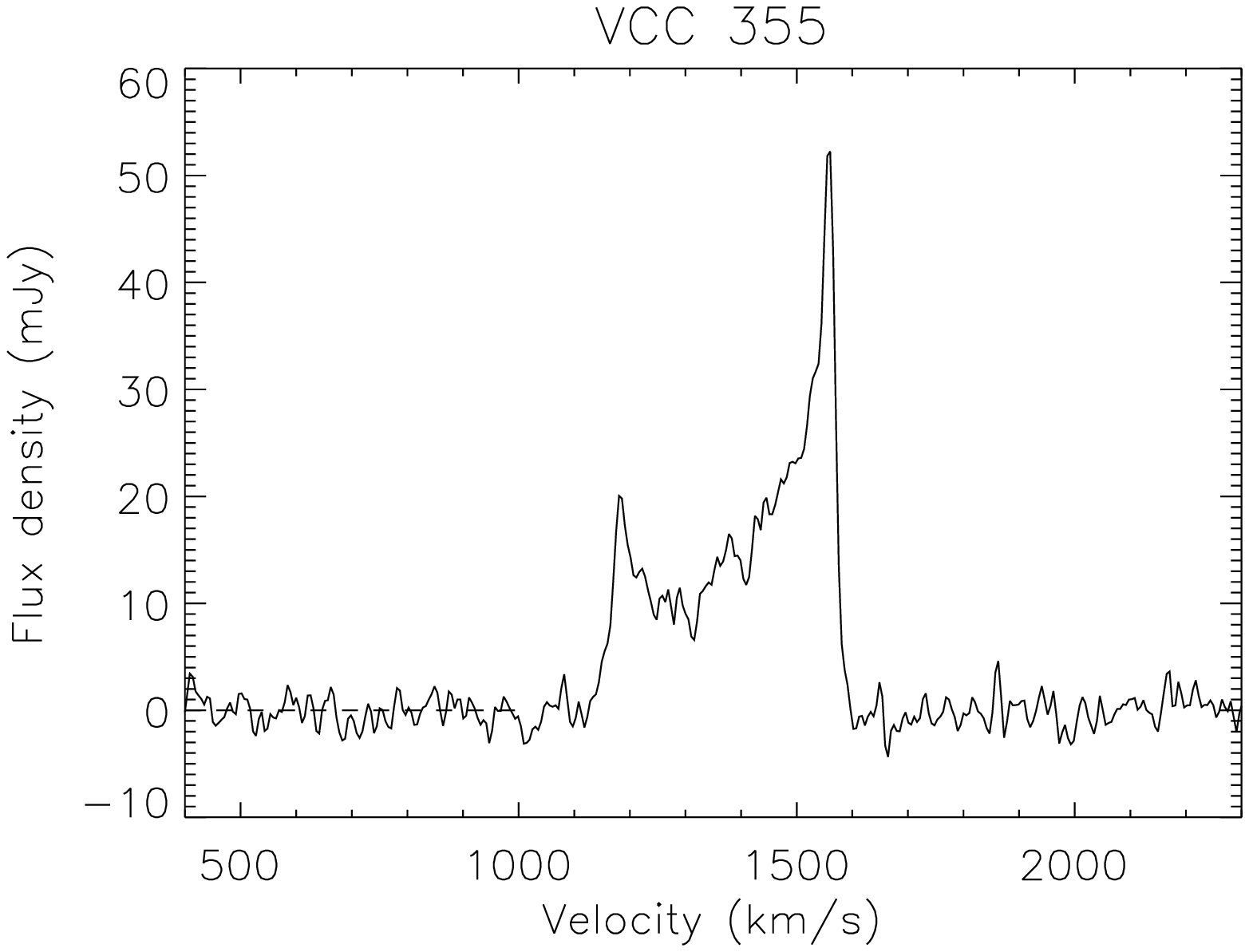}
\includegraphics[width=0.24\textwidth]{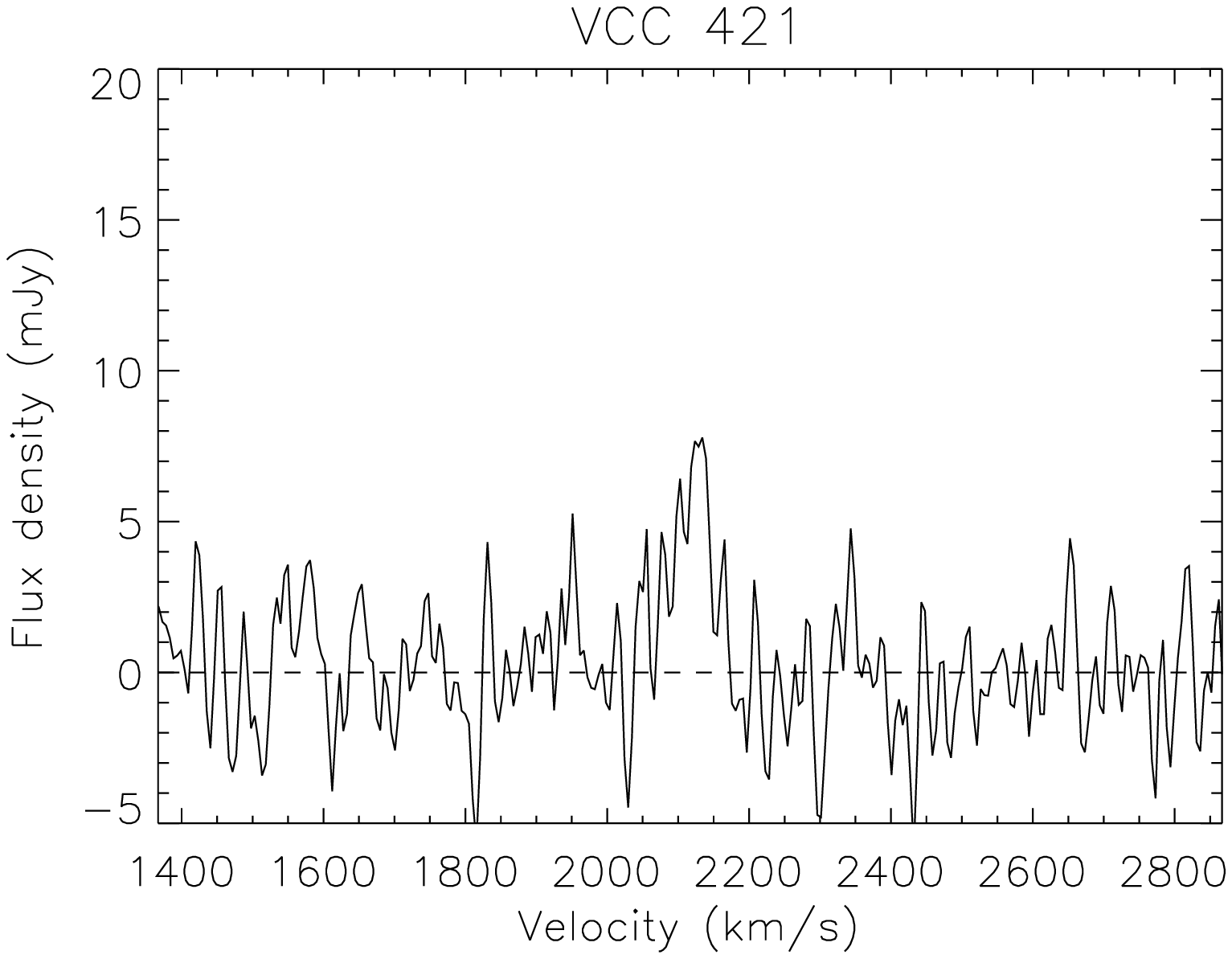}
\includegraphics[width=0.24\textwidth]{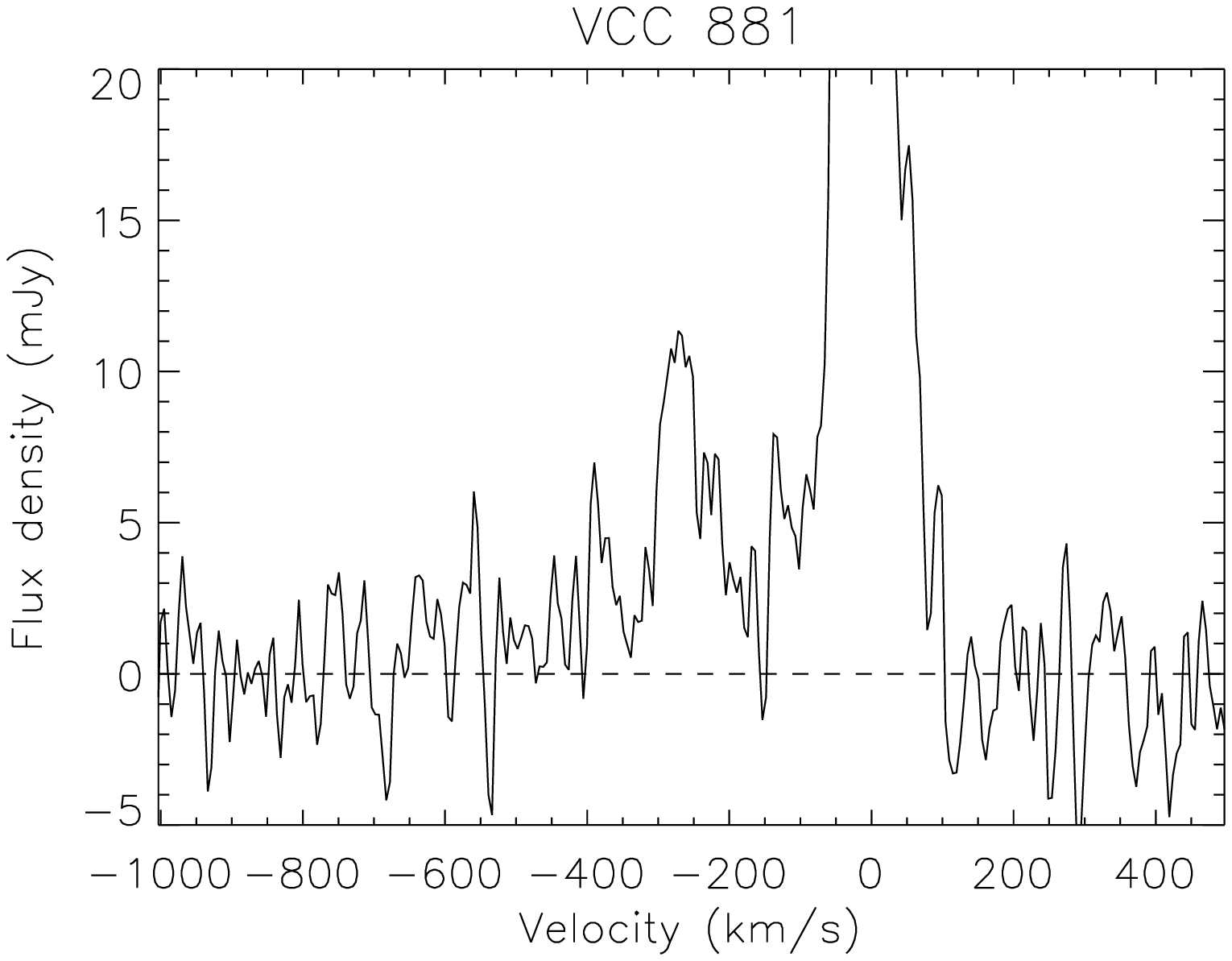}
\includegraphics[width=0.24\textwidth]{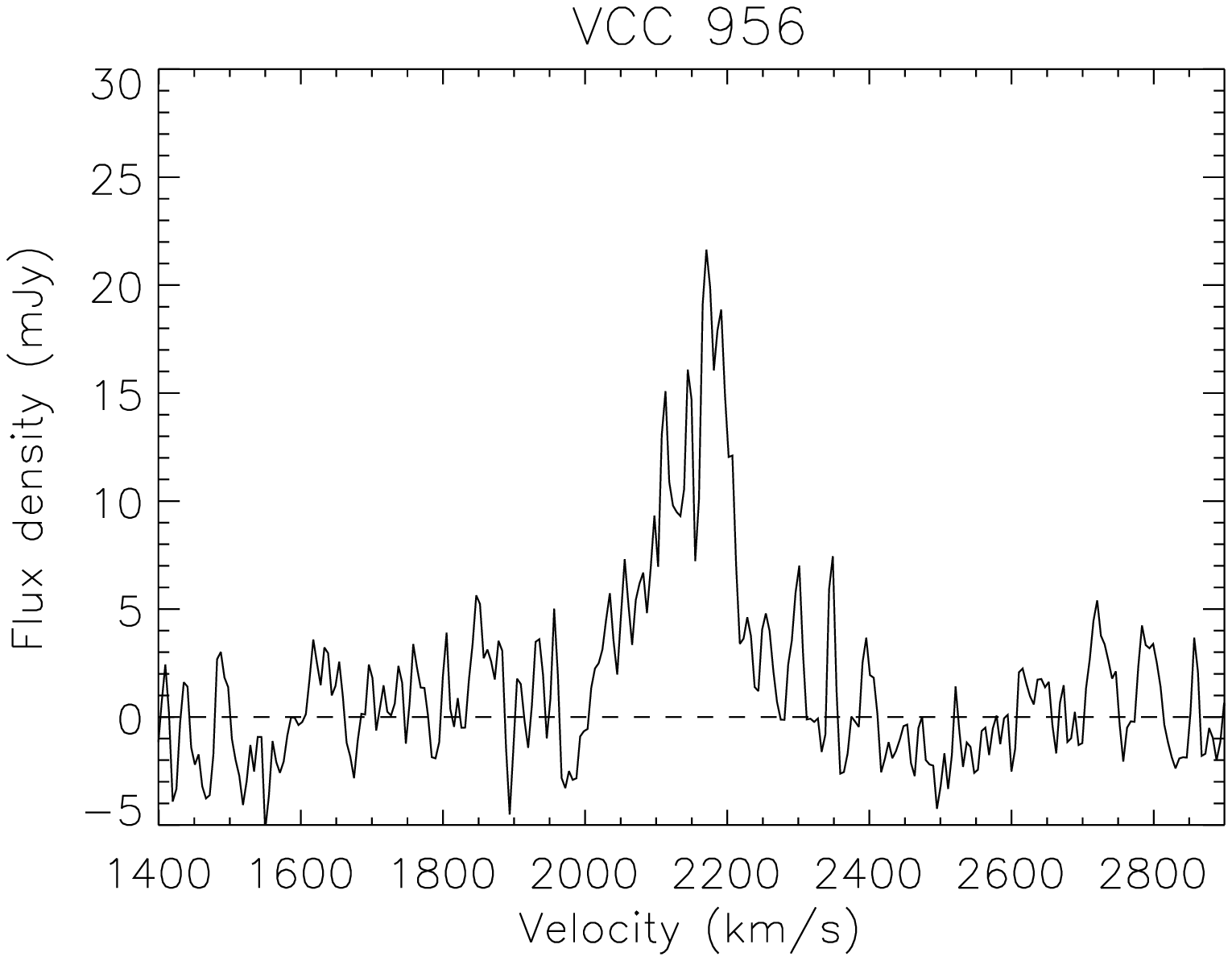}
\end{figure}
\begin{figure}[!ht]
\includegraphics[width=0.24\textwidth]{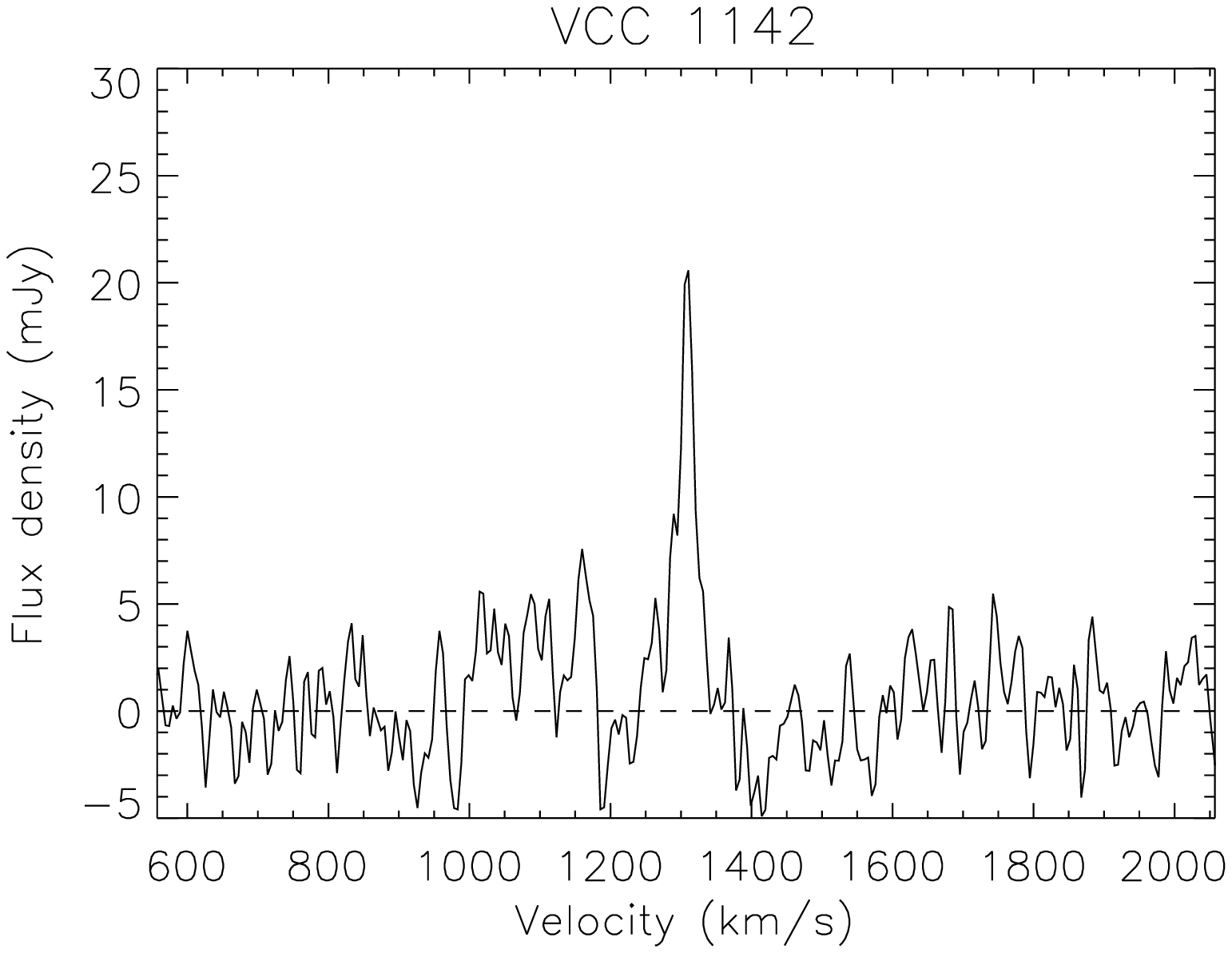}
\includegraphics[width=0.24\textwidth]{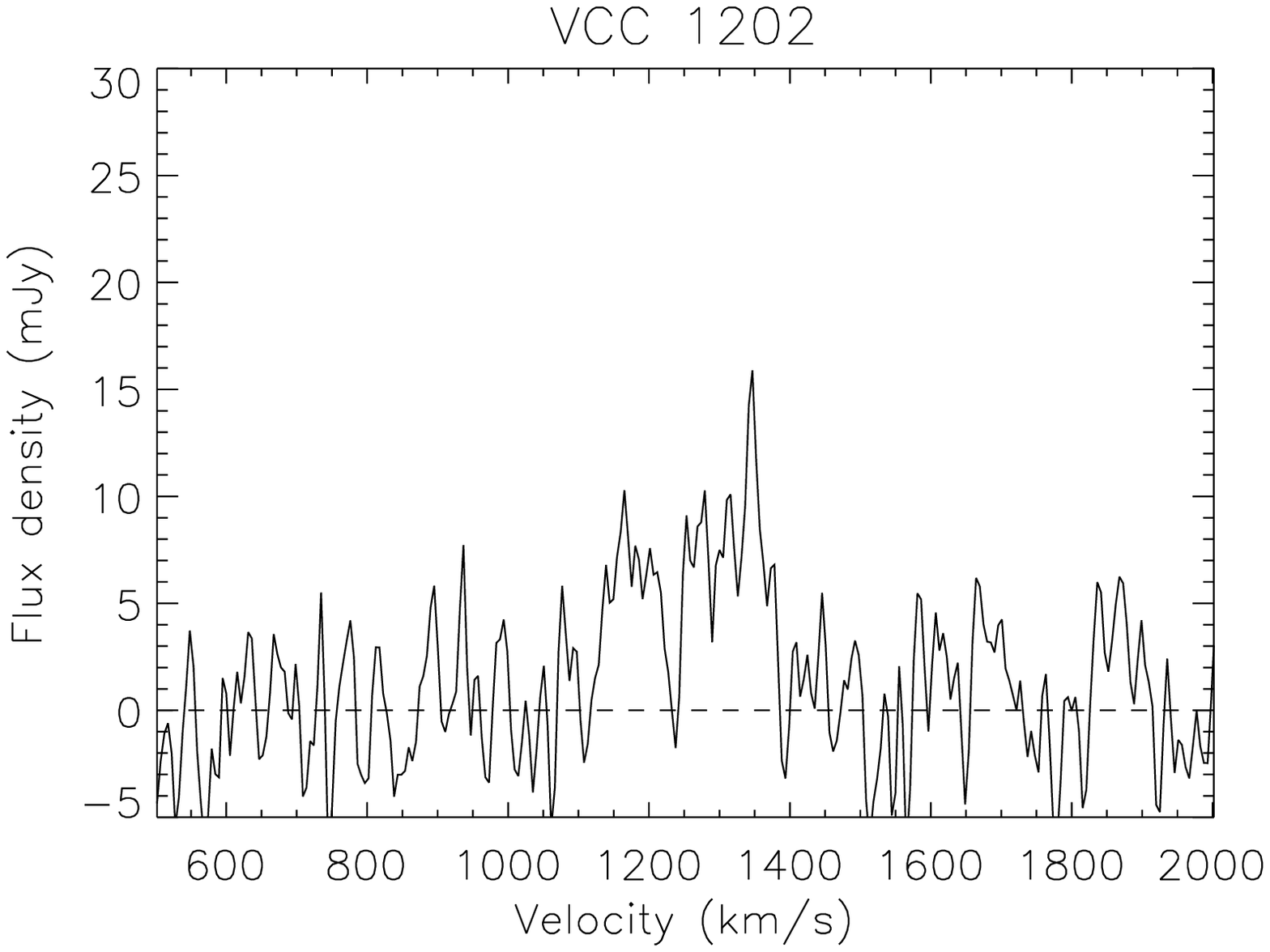}
\includegraphics[width=0.24\textwidth]{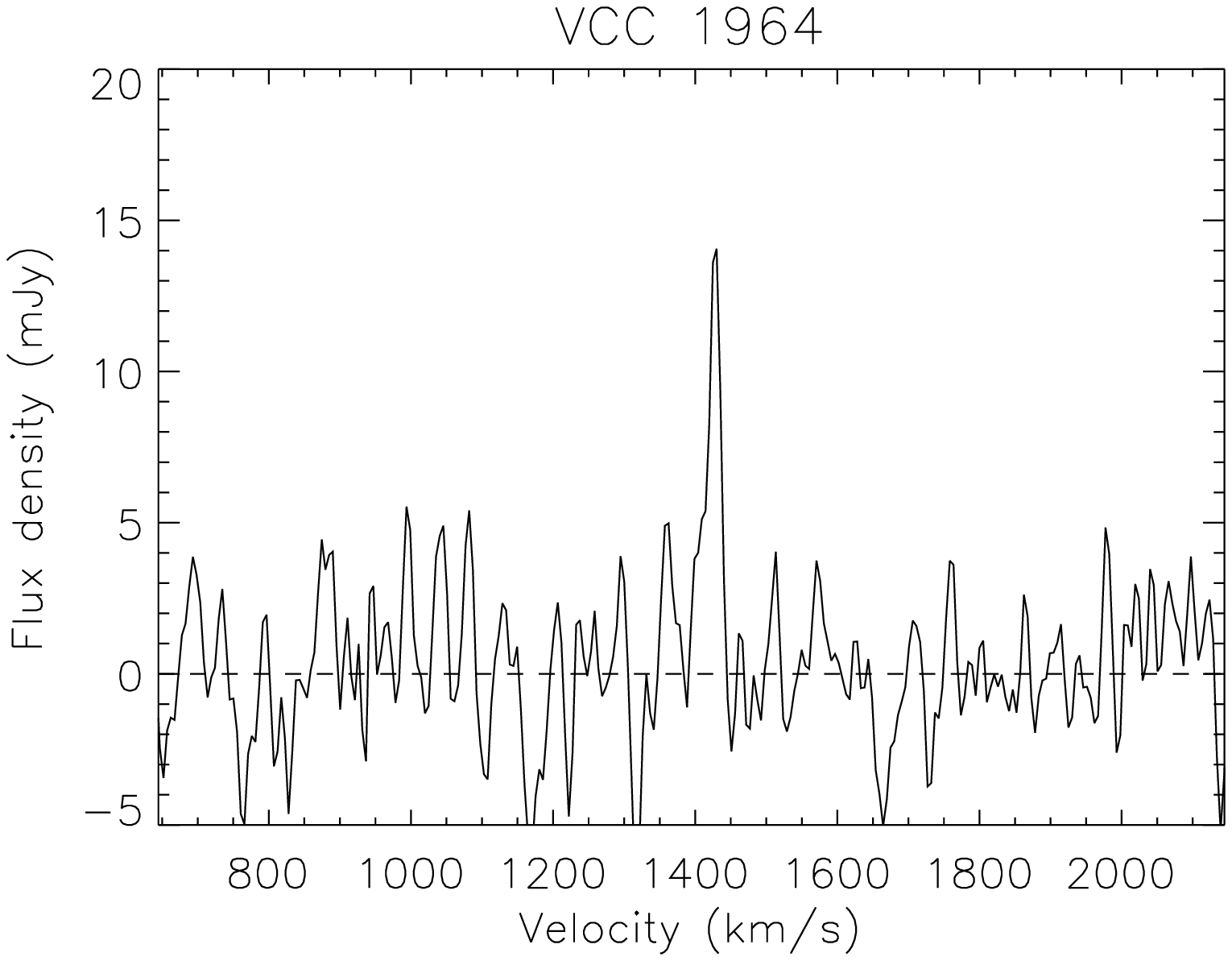}
\includegraphics[width=0.24\textwidth]{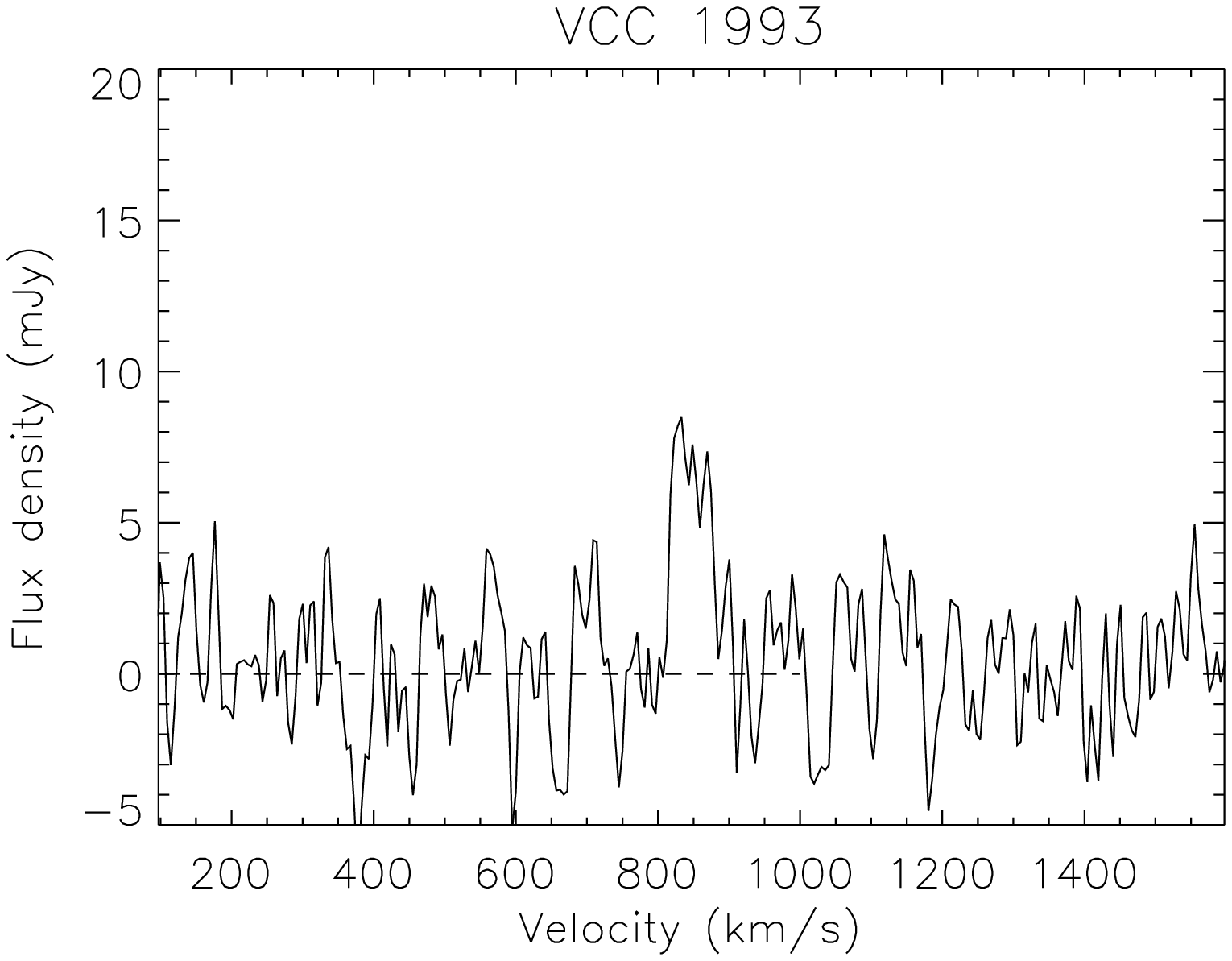}
\newline
\caption{The \hi spectra of the Virgo ETG detected by the ALFALFA
survey.}
\end{figure}

\section{The multi-phase interstellar medium of M86}

\begin{figure}
  \includegraphics[height=.3\textheight]{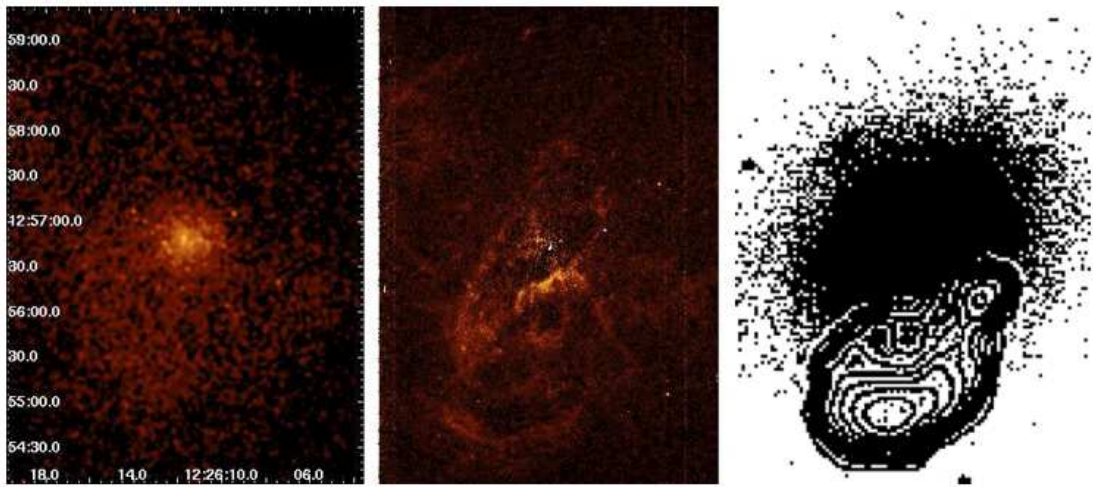}
  \caption{A comparison between the Chandra 0.5-2 keV (to the left), H$\alpha$
  (at the center, Trinchieri \& di Serego Alighieri 1991), and VLA 21-cm maps (to
  the right, Li \& van Gorkom 2001) of M86 (VCC
   881). All images have the same scale and cover the same area of the
   sky.}
\end{figure}

The brightest ETG detected in \hi in the Virgo cluster is M86 (VCC 881,
NGC 4406), an S0/E3 galaxy with a long plume of X-ray emission to the
North-West, which is thought to be due to hot gas swept back by the ram
pressure caused by the motion of M86 through the intracluster medium
(Forman et al. 1979). M86 is also the ETG with the largest $H\alpha$
luminosity found in the survey of ionised gas in a sample of ETG with hot
gas by Trinchieri \& di Serego Alighieri (1991). The rings and filaments
visible in their $H\alpha$ image, reproduced at the center of Fig. 2, suggest that
the gas might result from a recent interaction with a gas rich object.
This possibility is reinforced by an estimate of the evaporation time for
a cold gas cloud embedded in the hot ISM of M86, which is of the order of 1 Gyr.
Therefore cold gas cannot survive for very long in M86 and, if present, must have
been accreted less than a gigayear ago. 

A candidate gas-donor could be NGC
4406B (VCC 882), a dE galaxy located at a projected distance of 1.4 arcmin
to the North-East of the centre of M86. This galaxy is also located at
one end of a $\sim$28 kpc long dust trail in the halo of M86, which is
thought to be stripped material from the dwarf (Elmegreen et al. 2000).
A less likely candidate gas-donor could be the Virgo spiral galaxy
NGC 4388, which is thought to be the origin of the 110 kpc long plume of
\hi by
stripping in the hot intracluster medium (Oosterloo \& van Gorkom, 2005).
This plume of \hi passes at a projected distance of 2 arcmin to the
South-East of M86, but has a radial velocity of about 2200 km/s, very
different from the -280 km/s measured with the VLA for the \hi cloud
coincident with the ionised gas (to the right in Fig. 2, Li \& van Gorkom, 2001) and from
the \hi velocity measured from ALFALFA for M86 (-302 km/s).

\section{Conclusions}

Only very few ETG in the Virgo cluster (2.3\%) contain neutral hydrogen.
The relatively short evaporation time for the cold gas
embedded in a hot halo is a good reason for the scarcity of \hi in the
bright ETG with an X-ray halo. The exceptions, like M86, show evidence
for an interaction with a companion, which could have recently supplied
the cold gas. The few dwarf ETG detected in \hi are in fact at the border of
the ETG classification and could be later type galaxies, for which part
of the gas has been stripped away by the hot intracluster medium.



\begin{thebibliography}{9}

\bibitem{Bin85} B. Binggeli, A. Sandage \& G.~A. Tammann, \emph{AJ}
\textbf{90}, 1681 (1985).

\bibitem{diS07} S. di Serego Alighieri, G. Gavazzi, C. Giovanardi, M.
Grossi, M.~P. Haynes, B.~R. Kent, R.~A. Koopmann, S. Pellegrini, M.
Scodeggio \& G. Trinchieri, \emph{A\&A} \textbf{474}, 851 (2007).

\bibitem{Elm00} D.~M. Elmegreen, B.~G. Elmegreen, F.~R. Chromey \& M.~S.
Fine, \emph{AJ} \textbf{120}, 733 (2000).

\bibitem{For79} W. Forman, J. Schwarz, C. Jones, W. Liller \& A.~C.
Fabian, \emph{ApJ} \textbf{234}, L27 (1979).

\bibitem{Gio07} R. Giovanelli, M.P. Haynes, B.R. Kent, A. Saintonge, S.
Stierwalt, A. Altaf, T. Balonek, N. Brosch, S. Brown \& B. Catinella, \emph{AJ}
\textbf{133}, 2569 (2007).

\bibitem{Li01} Y. Li \& J.~H. van Gorkom, \emph{ASP Conf. Series}, Vol.
240, p. 637 (2001).

\bibitem{Oos05} T. Oosterloo \& J.~H. van Gorkom, \emph{A\&A}
\textbf{437}, L19 (2005).

\bibitem{Tri91} G. Trinchieri \& S. di Serego Alighieri, \emph{AJ}
\textbf{101}, 1647 (1991).

\end{thebibliography}
\end{document}